\documentclass[12pt,a4paper]{article}
\usepackage[margin=1in]{geometry}
\usepackage[T1]{fontenc}\usepackage{lmodern,microtype}
\usepackage{amsmath,amssymb,booktabs,graphicx,xcolor,tikz,hyperref,multicol}
\usepackage{caption}
\usepackage{titlesec}
\titlespacing*{\section}{0pt}{2.0ex plus .6ex minus .3ex}{.9ex plus .2ex}
\titlespacing*{\subsection}{0pt}{1.5ex plus .4ex minus .2ex}{.6ex plus .2ex}
\usepackage[section]{placeins}
\usetikzlibrary{arrows.meta,calc,positioning}
\hypersetup{colorlinks=true,linkcolor=blue!55!black,urlcolor=blue!55!black,citecolor=blue!55!black}
\newcommand{\R}{\mathbb R}\newcommand{\Sph}{\mathbb S}
\title{\bfseries Directional Statistics on Circles and Spheres\\[4pt]\large\itshape A New Angle on Data}
\author{Siddhesh Kulkarni\thanks{Corresponding author: \href{mailto:siddhesh.s.kulkarni@gmail.com}{siddhesh.s.kulkarni@gmail.com}}\\[4pt]
\normalsize Bristol Myers Squibb, 5 Giralda Farms, Madison, NJ 07940, USA}
\date{}
\begin{document}\maketitle
\section*{Summary}
Two compass bearings near north can have an ordinary arithmetic mean pointing south: the difficulty is that angles live on a circle, not a straight line. Using real wind observations, diffusion-MRI orientations, and geometric examples from machine learning, this article introduces circular and axial averages, directional variability, and probability models on circles and spheres. The guiding principle is to identify the geometry and symmetries of the observations before choosing a statistical method. \par
\noindent\textbf{Keywords:} directional statistics; circular statistics; spherical data; axial data; geometry-aware statistics; von Mises distribution; von Mises--Fisher distribution; Watson distribution; cosine similarity; machine learning; medical imaging.

\section{A statistical puzzle on the compass}
Suppose a ship records two headings, $1^\circ$ and $359^\circ$, measured clockwise from north. Both indicate almost due north. An arithmetic average gives
\begin{equation}\frac{1^\circ+359^\circ}{2}=180^\circ,\end{equation}
which points due south! The numbers are correct, but they have been treated as if they lay on a straight number line. On a compass, $0^\circ$ and $360^\circ$ are the same direction. The labels wrap around.

This distinction is not confined to navigation. Wind and ocean-current directions, the timing of daily biological cycles, animal migration bearings, cell and fiber orientations, geological directions, robot headings, surface normals in computer vision, and normalized feature vectors can all require similar care. The broader field of \emph{directional statistics} studies observations that lie on circles, spheres, and related geometric spaces \cite{mardiajupp,pewsey}. Circular methods are already used across biology, including animal navigation and chronobiology \cite{landler}. Our theme is not that ordinary statistics is ``wrong'', but that it quietly assumes a geometry. Familiar notions such as averages, spread, distance, probability, and similarity must sometimes be reinterpreted when observations wrap around, live on a sphere, or identify $\mathbf x$ with $-\mathbf x$. The central question is therefore: \emph{what should routine statistical ideas mean on a non-Euclidean sample space?}

\section{Where do directional data appear?}
Directional statistics is not a niche subject tied to compasses. The same geometric issues recur whenever a variable wraps around, lies on a sphere, represents an unoriented axis, or is deliberately normalized to unit length. Table~\ref{tab:applications} gives representative examples. Its purpose is breadth rather than completeness: the scientific meaning of a direction matters as much as its mathematical representation.

\begin{table}[!htbp]\centering\small
\begin{tabular}{@{}p{.20\linewidth}p{.36\linewidth}p{.33\linewidth}@{}}\toprule
\textbf{Area} & \textbf{Example} & \textbf{Geometric issue}\\\midrule
Weather / ocean & Wind, current, wave direction & Circular wraparound; multimodality\\
Ecology / chronobiology & Migration bearings; circadian phase & Preferred direction; 24-hour periodicity\\
Microscopy / medicine & Cell, collagen, or fiber orientation & Often axial: $\theta\equiv\theta+180^\circ$\\
Geology / astronomy & Faults, paleomagnetism, sky directions & Circular, spherical, or axial\\
Robotics / computer vision & Heading, attitude, surface normals & Periodic angles and 3-D orientation\\
Machine learning / AI & Unit-normalized features and embeddings & High-dimensional sphere; angular similarity\\\bottomrule
\end{tabular}
\caption{Directional data across science and technology. The appropriate analysis depends on geometry, symmetry, and the scientific meaning of the measurement; representative applications are discussed in standard directional-statistics, biology, diffusion-MRI, and spherical-clustering references \cite{mardiajupp,landler,jones,banerjee}.}\label{tab:applications}
\end{table}

Three examples make the recurring patterns concrete. In circadian biology, 11:55 PM and 12:05 AM are ten minutes apart, not almost 24 hours apart. In microscopy, an elongated cell at $30^\circ$ may have the same physical orientation as one recorded at $210^\circ$ when its ends are indistinguishable. In machine learning, normalizing a vector makes angular comparison natural but deliberately removes information about overall magnitude. Although these problems involve periodicity, axial symmetry, and spherical similarity, they share one lesson: \emph{identify the sample space before choosing the statistic.}

\section{What kind of directional problem do I have?}
\begin{figure}[!htbp]\centering
\includegraphics[width=.98\linewidth]{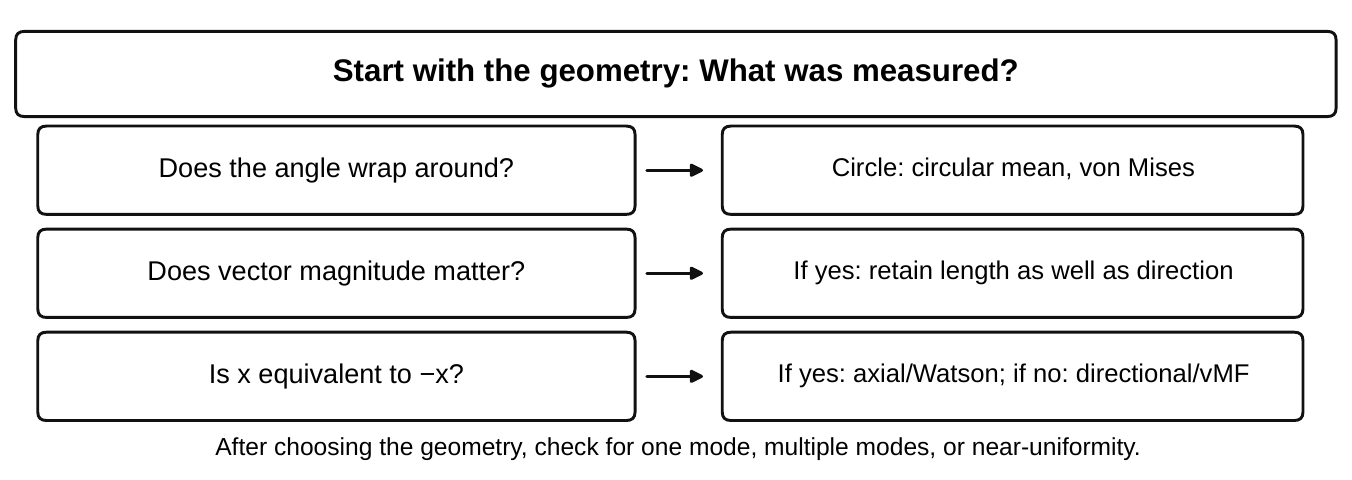}
\caption{A geometry-first decision guide. First check periodicity, then whether vector magnitude matters, and finally whether opposite vectors describe the same physical state. These are diagnostic questions, not a unique model-selection rule.}\label{fig:decision}
\end{figure}

The decision in Figure~\ref{fig:decision} separates several questions that are easily conflated. Is the variable intrinsically periodic? Does magnitude carry scientific information? If we use a unit vector, are $\mathbf x$ and $-\mathbf x$ different states or two labels for the same orientation? Only after answering those questions should we choose a summary or probability model.

\section{A real wind example: the same north, two different arithmetic means}
\begin{figure}[!htbp]\centering
\includegraphics[width=.98\linewidth]{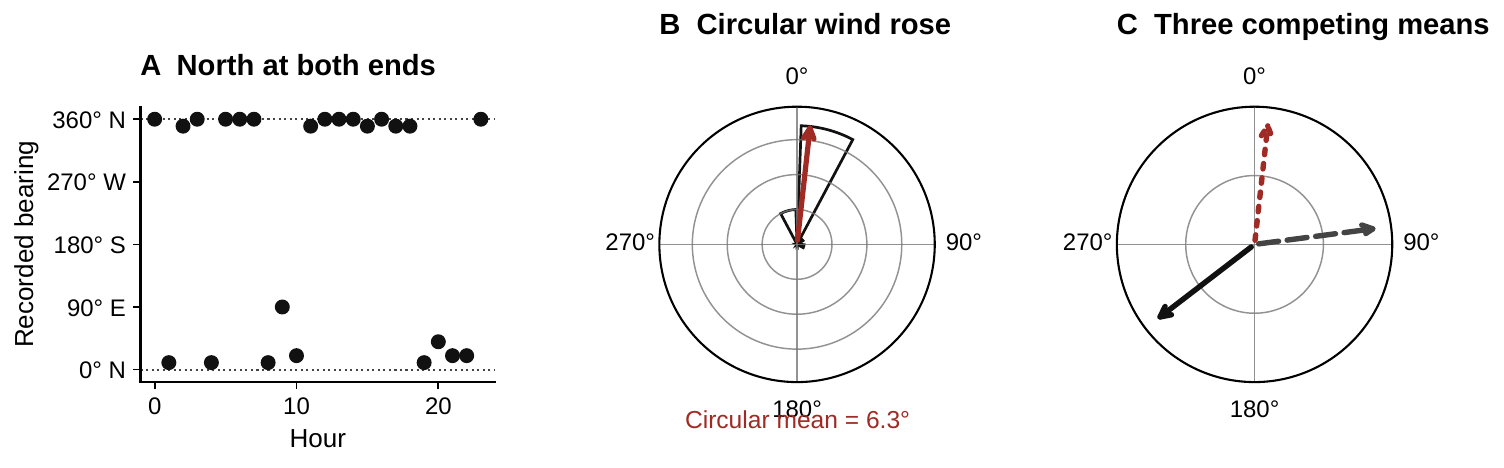}
\caption{\textbf{Real hourly wind directions at JFK Airport on 15 January 2013.} (A) A straight scale places the two equivalent labels for north, $0^\circ$ and $360^\circ$, at opposite ends. (B) An outlined wind rose restores their adjacency; the circular mean is about $6.3^\circ$ and $\bar R=0.94$. (C) Replacing the label $360^\circ$ by the equivalent label $0^\circ$ changes the arithmetic mean from $232.5^\circ$ to $82.5^\circ$, while the circular mean remains unchanged. In panel C, the solid, dashed, and dotted arrows represent $232.5^\circ$, $82.5^\circ$, and $6.3^\circ$, respectively. Data: \texttt{nycflights13::}\allowbreak\texttt{weather} \cite{nycflights13}.}\label{fig:wind}
\end{figure}

Figure~\ref{fig:wind} uses 24 hourly wind directions recorded at JFK Airport on 15 January 2013 in the \texttt{nycflights13::}\allowbreak\texttt{weather} data, derived from ASOS observations obtained through the Iowa Environmental Mesonet \cite{nycflights13}. Meteorological wind direction records where the wind comes from. The point of this one-day example is simple: north is the same physical direction whether we write it as $0^\circ$ or $360^\circ$.

With north recorded as $360^\circ$, the ordinary arithmetic mean is $232.5^\circ$. If we change only the numerical label $360^\circ$ to the equivalent label $0^\circ$, the ordinary mean becomes $82.5^\circ$. Nothing about the weather has changed, yet the arithmetic mean moves by $150^\circ$. The circular mean does not depend on that arbitrary cut: it is $6.3^\circ$, just east of north, with mean resultant length $\bar R=0.94$. This is why a circular summary is needed. The 24 measurements are sequential hourly observations and may be serially dependent; here they are used to illustrate geometry and coding, not to make an independence-based inferential claim.

\section{What does an average look like on a circle?}
For convenience, let angles now be measured counterclockwise from the horizontal axis, in radians. The point corresponding to an angle $\theta$ is
\begin{equation}\mathbf x(\theta)=\begin{pmatrix}\cos\theta\\\sin\theta\end{pmatrix},\qquad \|\mathbf x(\theta)\|=1.\label{eq:circle}\end{equation}
The unit circle is the set $\Sph^1=\{\mathbf x\in\R^2:\|\mathbf x\|=1\}$. The key property is $\mathbf x(\theta+2\pi)=\mathbf x(\theta)$: the same point has many equivalent numerical labels.

\subsection{The mean direction as an average of arrows}
The unfamiliar-looking formulas for a circular mean come from an ordinary operation: averaging vectors coordinate by coordinate. An angle $\theta_i$ corresponds to the unit vector
\[
\mathbf x_i=(\cos\theta_i,\sin\theta_i).
\]
Its cosine is the horizontal coordinate and its sine is the vertical coordinate. Therefore the average horizontal and vertical coordinates are
\begin{equation}
\bar C=\frac1n\sum_{i=1}^n\cos\theta_i,\qquad
\bar S=\frac1n\sum_{i=1}^n\sin\theta_i.\label{eq:cs}
\end{equation}
Thus $(\bar C,\bar S)$ is simply the average arrow. For example, the directions $0^\circ$ and $90^\circ$ correspond to $(1,0)$ and $(0,1)$, whose average is $(1/2,1/2)$; that average arrow points at $45^\circ$.

We now need to convert the coordinates of the average arrow back into an angle. Recall that for a right triangle
\[
\tan\theta=\frac{\text{vertical component}}{\text{horizontal component}}.
\]
The inverse tangent, written $\arctan$, reverses this relationship: $\arctan(u)=\theta$ when $\tan\theta=u$ on the principal interval $(-\pi/2,\pi/2)$. Using only $\arctan(y/x)$ is not enough for a full circle, however, because opposite quadrants can have the same ratio $y/x$. The two-coordinate angle function $\operatorname{atan2}(y,x)$ uses the signs of both coordinates to return the correct direction of the vector $(x,y)$.

Hence the circular mean is simply the angle of the average vector:
\begin{equation}
\bar\theta=\operatorname{atan2}(\bar S,\bar C).\label{eq:mean}
\end{equation}
For $1^\circ$ and $359^\circ$, the sine components nearly cancel while the cosine components reinforce one another, so $(\bar C,\bar S)$ points almost exactly north. The apparent special formula is therefore just ``average the arrows, then read off the direction of the average arrow.''

There is also a geometric way to understand this rule. For a proposed direction $\phi$, consider the sum of squared distances between the observed unit vectors and the proposed unit vector:
\begin{align}
\sum_{i=1}^n\|\mathbf x(\theta_i)-\mathbf x(\phi)\|^2
&=\sum_{i=1}^n\bigl(2-2\cos(\theta_i-\phi)\bigr)\\
&=2n-2n\bigl(\bar C\cos\phi+\bar S\sin\phi\bigr).\label{eq:least}
\end{align}
Thus minimizing squared \emph{chord} distances is equivalent to choosing the direction that maximizes the dot product with the average vector. This is precisely Equation~\eqref{eq:mean}. It is one principled notion of a circular center, though other notions (for example, based on shortest arc distances) are also possible.

\subsection{When an average should not exist}
Take $0^\circ$ and $180^\circ$. Their vectors are $(1,0)$ and $(-1,0)$, which sum to zero. The circular mean direction is then \emph{undefined}. That is an informative result: these two directions do not single out a unique preferred heading. Software should not silently report a direction in this case.

\section{What does variability mean for directions?}
The length of the average vector is called the \emph{mean resultant length}:
\begin{equation}\bar R=\sqrt{\bar C^2+\bar S^2},\qquad 0\leq\bar R\leq1.\label{eq:r}\end{equation}
If every arrow points in exactly the same direction, $\bar R=1$. If the arrows cancel, $\bar R=0$. A common analogue of variance is the \emph{circular variance}
\begin{equation}V_{\rm circ}=1-\bar R,\end{equation}
and one commonly used circular standard deviation is $s_{\rm circ}=\sqrt{-2\log\bar R}$ when $\bar R>0$. These quantities summarize concentration, but they do not describe shape completely: two tight clusters pointing in opposite directions can have $\bar R\approx0$. Standard circular inference therefore also includes tools such as the Rayleigh test for departure from uniformity and, under appropriate assumptions, the Watson--Williams test for comparing mean directions \cite{mardiajupp,landler}.

\section{What does a probability model look like on a circle?}
How can we describe random variation around a preferred direction? On the real line, a normal distribution puts greater probability near its center. For circular data, a natural counterpart is the \emph{von Mises distribution}:
\begin{equation}
f(\theta\mid\mu,\kappa)=\frac{\exp\{\kappa\cos(\theta-\mu)\}}{2\pi I_0(\kappa)},\quad 0\leq\theta<2\pi.\label{eq:vm}
\end{equation}
Here $\mu$ is the preferred direction, $\kappa\geq0$ is a \emph{concentration parameter}, and $I_0$ is a modified Bessel function. Its role is normalization:
\begin{equation}I_0(\kappa)=\frac{1}{2\pi}\int_0^{2\pi}\exp\{\kappa\cos t\}\,dt.\label{eq:bessel}\end{equation}
This integral makes the total area under the circular density equal to one. Readers need not evaluate it by hand.

Why the cosine? The identity $\cos(\theta-\mu)=\mathbf x(\theta)^\top\mathbf x(\mu)$ measures alignment. It is largest when the observed direction matches the preferred direction and smallest when it points the other way. When $\kappa=0$, Equation~\eqref{eq:vm} is uniform around the circle. As $\kappa$ increases, the distribution concentrates near $\mu$. For large $\kappa$, a small-angle approximation $\cos\delta\approx1-\delta^2/2$ gives
\begin{equation}\exp\{\kappa\cos\delta\}\approx e^\kappa\exp\{-\kappa\delta^2/2\}.\end{equation}
Thus, near its peak, the \emph{kernel} of a highly concentrated von Mises density resembles that of a normal density with angular variance approximately $1/\kappa$; this local approximation suppresses the changing normalizing constant. It is not a claim that a normal distribution automatically respects circular wraparound.

\subsection{Why the circular mean reappears in likelihood}
Suppose $\theta_1,\ldots,\theta_n$ are independent von Mises observations. If $\kappa>0$ is fixed, the part of their log-likelihood that depends on $\mu$ is
\begin{equation}\ell(\mu)=\kappa\sum_{i=1}^n\cos(\theta_i-\mu)+\text{constant}.\end{equation}
Using the same trigonometric expansion as before, maximizing $\ell(\mu)$ selects the direction of $(\bar C,\bar S)$. Hence the circular mean is also a maximum-likelihood estimate of the preferred direction under this model, whenever the resultant is nonzero. The geometry and the statistical model lead to the same answer.

\begin{figure}[!htbp]\centering
\includegraphics[width=.94\linewidth]{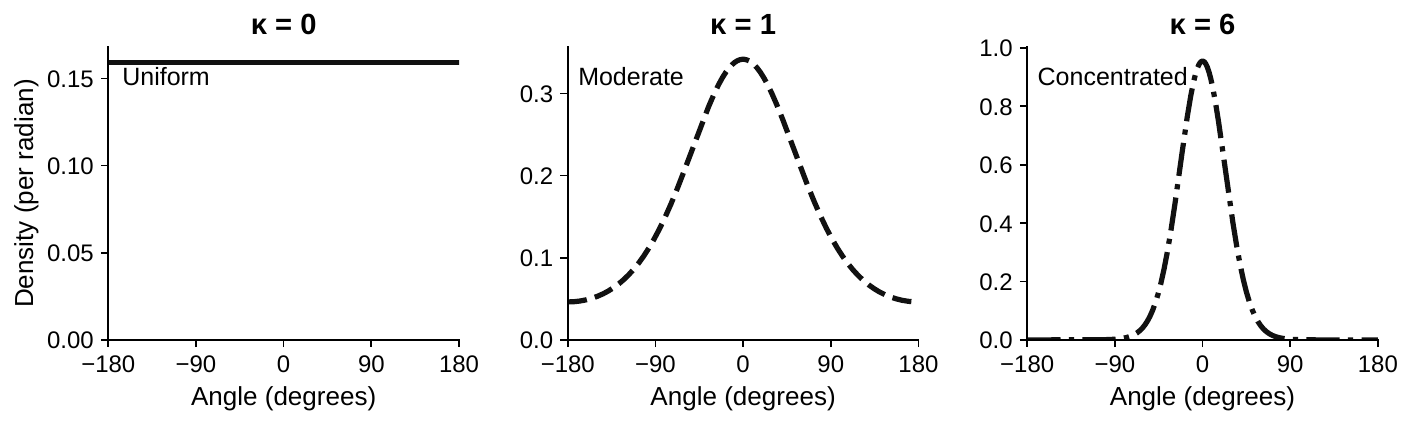}
\caption{Properly normalized von Mises densities. Increasing concentration $\kappa$ sharpens the peak around zero.}\label{fig:vm}
\end{figure}

\section{From a circle to a sphere: the same idea in higher dimensions}
A compass bearing needs one angle. A three-dimensional direction needs more information, but the geometric idea survives. A unit vector in $p$ dimensions belongs to
\begin{equation}\Sph^{p-1}=\{\mathbf x\in\R^p:\|\mathbf x\|=1\}.\end{equation}
The circle is $\Sph^1$ and the familiar sphere is $\Sph^2$. The \emph{von Mises--Fisher} (vMF) distribution extends the von Mises idea:
\begin{equation}f(\mathbf x\mid\boldsymbol\mu,\kappa)=C_p(\kappa)\exp\{\kappa\boldsymbol\mu^\top\mathbf x\},\quad \mathbf x,\boldsymbol\mu\in\Sph^{p-1}.\label{eq:vmf}\end{equation}
The vector $\boldsymbol\mu$ gives the preferred direction and $\kappa\geq0$ controls concentration. $C_p(\kappa)$ is the normalizing constant for surface area on the sphere; explicitly, for $\kappa>0$,
\begin{equation}C_p(\kappa)=\frac{\kappa^{p/2-1}}{(2\pi)^{p/2}I_{p/2-1}(\kappa)}.\end{equation}
Here $I_\nu$ is the modified Bessel function of order $\nu$. At $\kappa=0$, the distribution is uniform on the sphere, with the corresponding limiting normalizing constant.

The apparent new formula is really the same old idea. For unit vectors, $\boldsymbol\mu^\top\mathbf x=\cos\alpha$, where $\alpha$ is the angle between them. On $\Sph^1$, this becomes $\cos(\theta-\mu)$, exactly the quantity in Equation~\eqref{eq:vm}. For independent vMF observations with nonzero vector sum, the maximum-likelihood preferred direction is the normalized sum $\sum_i\mathbf x_i/\|\sum_i\mathbf x_i\|$. Thus averaging arrows extends naturally from two to many dimensions \cite{mardiajupp,sra}. In high dimensions, however, evaluating the vMF normalizing constant and estimating $\kappa$ can become numerically important; practical approximations are discussed by Banerjee et al. \cite{banerjee}.

\section{When opposite directions mean the same thing}
Imagine a compass needle pointing north. Reversing it to south changes its heading. Now imagine a straight, unoriented fiber running north--south. Reversing the arrow used to draw the fiber does \emph{not} change the fiber's physical axis. In the second situation, $\mathbf x$ and $-\mathbf x$ are two descriptions of the same observation. Such data are called \emph{axial} or \emph{unoriented}.

A classical axial model is the \emph{Watson distribution}, whose density (relative to uniform surface measure) is proportional to
\begin{equation}f(\mathbf x\mid\boldsymbol\mu,\kappa)\propto\exp\{\kappa(\boldsymbol\mu^\top\mathbf x)^2\}.\label{eq:watson}\end{equation}
The square ensures that $f(\mathbf x)=f(-\mathbf x)$. When $\kappa>0$, observations tend to align with the axis $\pm\boldsymbol\mu$. When $\kappa<0$, the density favors directions perpendicular to that axis. The vMF model, by contrast, distinguishes $\boldsymbol\mu$ from $-\boldsymbol\mu$ when $\kappa>0$.

This distinction arises naturally in diffusion magnetic resonance imaging (MRI). In a single-tensor model the diffusion-weighted signal has the form $S(\mathbf g,b)/S_0=\exp\{-b\mathbf g^\top\mathbf D\mathbf g\}$, where $\mathbf g$ is a measurement direction and $\mathbf D$ is the fitted diffusion tensor. Replacing $\mathbf g$ by $-\mathbf g$ leaves the quadratic form unchanged, reflecting antipodal symmetry. The principal eigenvector of $\mathbf D$ therefore represents an \emph{axis}; its sign is not an anatomical label \cite{basser,jones,garyfallidis}. A statistical model that changed because software happened to return the opposite eigenvector sign would be scientifically inappropriate.

\subsection{A medical example: what do diffusion-MRI data look like?}
\begin{figure}[!htbp]\centering
\includegraphics[width=.98\linewidth]{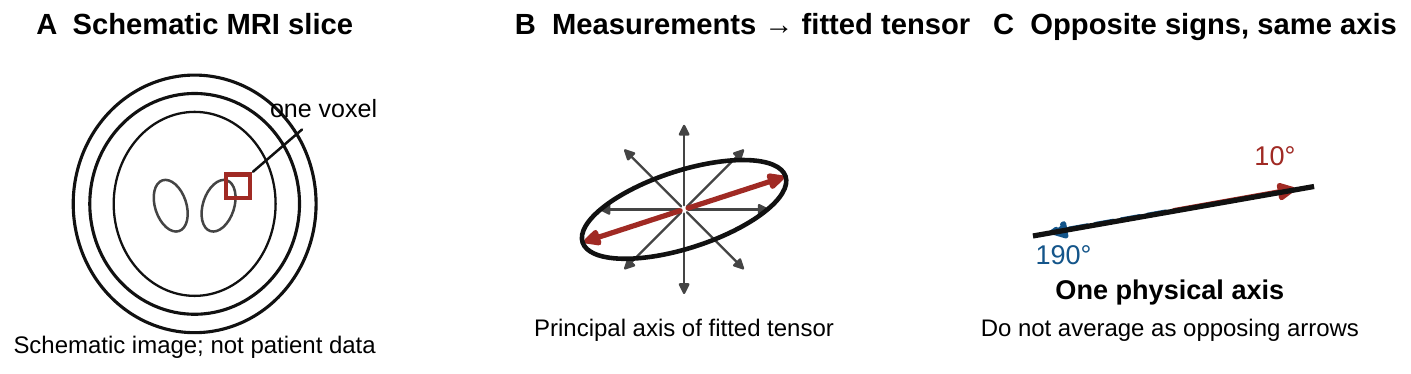}
\caption{\textbf{From an MRI image to an axial statistical problem.} (A) A vector-drawn, schematic MRI-like slice illustrates the voxelized image on which diffusion measurements are made; diffusion MRI acquires many volumes under different gradient directions. (B) Within one voxel, the directional signal measurements are used to fit a local diffusion tensor, whose long principal axis summarizes orientation. (C) The orientations $10^\circ$ and $190^\circ$ describe the same physical axis; treating antipodal observations as ordinary arrows can therefore cause artificial cancellation. All panels are schematic rather than patient data \cite{jones,garyfallidis}.}\label{fig:medical}
\end{figure}

The phrase ``fiber orientation'' can sound abstract if one has never seen MRI data. Figure~\ref{fig:medical} therefore begins one step earlier. A diffusion-MRI experiment produces a series of grayscale brain volumes, each acquired with diffusion weighting applied along a specified gradient direction. The raw data are signal intensities in voxels, not tiny arrows drawn inside the brain. A model is then fitted to the directional signal pattern within each voxel. In the classical diffusion-tensor model, the fitted tensor can be pictured as an ellipsoid; its long principal axis summarizes the locally preferred diffusion orientation \cite{jones,garyfallidis}.

That fitted axis creates the statistical issue. The eigenvector returned by an algorithm may be $\mathbf v$ or $-\mathbf v$, but both describe the same physical line. Panel C isolates the consequence using six synthetic orientations,
$10^\circ,190^\circ,12^\circ,192^\circ,8^\circ,188^\circ$.
If they are treated as directed arrows, opposite signs cancel. If they are treated as axes, they represent one common orientation, $10^\circ$ modulo $180^\circ$.

A convenient estimator uses the $2\times2$ matrix
\begin{equation}
\mathbf M=\frac1n\sum_{i=1}^n\mathbf x_i\mathbf x_i^\top.
\label{eq:orientationmatrix}
\end{equation}
Replacing $\mathbf x_i$ by $-\mathbf x_i$ leaves $\mathbf M$ unchanged, because $(-\mathbf x_i)(-\mathbf x_i)^\top=\mathbf x_i\mathbf x_i^\top$. The eigenvector corresponding to the largest eigenvalue gives the dominant \emph{axis} (its sign is arbitrary). In two dimensions, an equivalent classical trick is to double each axial angle, compute an ordinary circular mean on the doubled angles, and divide the result by two. For the six values above, both approaches give $10^\circ$ modulo $180^\circ$. In real diffusion MRI, estimating tissue orientation from diffusion-weighted signals requires substantially more sophisticated measurement and modeling; the toy example isolates the statistical issue of antipodal symmetry.

\section{What does similarity mean in machine learning?}
Machine-learning systems often represent documents, images, or other observations by vectors in a feature space. When their direction rather than their length is of interest, a vector $\mathbf z\ne0$ can be normalized to $\mathbf x=\mathbf z/\|\mathbf z\|$. A common comparison is \emph{cosine similarity}:
\begin{equation}\operatorname{sim}(\mathbf z,\mathbf w)=\frac{\mathbf z^\top\mathbf w}{\|\mathbf z\|\,\|\mathbf w\|}.\label{eq:cos}\end{equation}
For normalized vectors, this is simply their dot product. Similar directions have similarity close to one, perpendicular vectors have similarity zero, and opposite directions have similarity minus one.

A useful identity follows by expanding a square:
\begin{equation}\|\mathbf x-\mathbf y\|^2=2-2\mathbf x^\top\mathbf y=2\{1-\operatorname{sim}(\mathbf x,\mathbf y)\},\qquad \|\mathbf x\|=\|\mathbf y\|=1.\label{eq:chord}\end{equation}
Thus, for unit vectors, chord distance and cosine similarity produce equivalent pairwise rankings. We should not claim that Euclidean distance is universally ``wrong'' on a sphere. The deeper issue is whether a particular probability model, clustering method, or scientific interpretation respects the underlying geometry. Directional distributions, including mixtures of vMF components, offer one principled approach to clustering directional observations \cite{sra,banerjee}.

There is an important caution. Normalizing a biomedical vector can discard meaningful magnitude information: a weak treatment response and a strong treatment response might have similar relative patterns but very different biological significance. Directional analysis is useful when relative pattern or orientation is the target of inference, not merely whenever normalization is mathematically possible.

\subsection{A machine-learning example: comparing three short texts}
Imagine that a text-retrieval system maps three phrases into two-dimensional vectors. For illustration, \emph{we choose these vectors by hand}; they are not outputs of a trained language model:
\begin{align*}
\text{A: ``cancer therapy''}&\longmapsto (1,0),\\
\text{B: ``tumor treatment''}&\longmapsto (0.8,0.6),\\
\text{C: ``weather forecast''}&\longmapsto (0,1).
\end{align*}
All three have length one. The cosine similarities to A are
\begin{equation}
\operatorname{sim}(\mathrm A,\mathrm B)=0.8,\qquad
\operatorname{sim}(\mathrm A,\mathrm C)=0.
\end{equation}
Thus B ranks above C as a match to A \emph{in this constructed feature space}. The example does not establish how any real embedding model would represent these phrases; actual embedding models typically operate in hundreds or thousands of dimensions. The two-dimensional picture is only a projection of the same angular idea.

Now multiply B by five. Its Euclidean length changes from $1$ to $5$, but its cosine similarity to A remains $0.8$. This makes cosine similarity useful when the \emph{pattern} of features matters more than the overall scale. In contrast, if vector magnitude encodes confidence, expression strength, or another scientifically important quantity, discarding it may be undesirable. Directional statistics therefore begins with a scientific decision about what should count as similar, not with an automatic normalization step.

\begin{figure}[!htbp]\centering
\includegraphics[width=.56\linewidth]{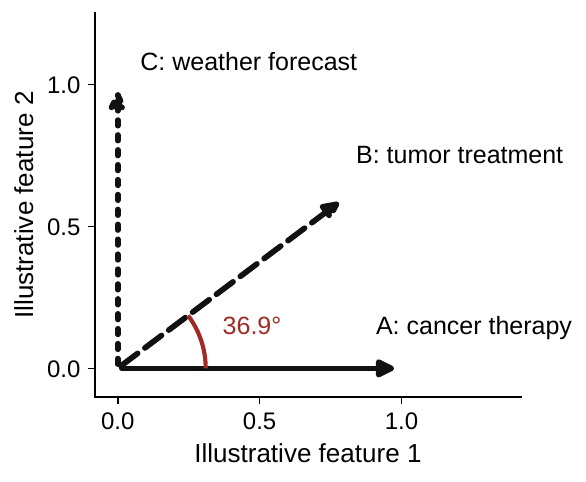}
\caption{An illustrative document embedding: cosine similarity between A and B is $0.8$, corresponding to an angle of approximately $36.9^\circ$. The vectors are constructed, not outputs of a trained model.}\label{fig:ml}
\end{figure}

\section{Explore it yourself: Directional Statistics Lab}
The companion \emph{Directional Statistics Lab} provides short, reproducible \textsf{R} scripts for readers and instructors to explore the article's main geometric ideas \cite{directionallab}. Version 0.1.0 contains five demonstrations: (1) circular mean, resultant length, and circular variance; (2) the observed JFK wind-direction example; (3) the effect of von Mises concentration $\kappa$; (4) axial orientation and the equivalence of angles separated by $180^\circ$; and (5) cosine similarity of normalized vectors. The examples generate figures and can be run separately or together using \texttt{R/run\_all.R}. A targeted suite of 38 numerical and edge-case checks passed for this initial release. This is a collection of teaching scripts, not yet an interactive application or a general-purpose \textsf{R} package.

The publicly accessible source is the \href{https://github.com/SiddheshKulkarni-source/directional-statistics-lab}{Directional Statistics Lab GitHub repository}. For exact reproducibility, the October 2026 v0.1.0 archive has its own DOI, \url{https://doi.org/10.5281/zenodo.23426629}. The project-level DOI, \url{https://doi.org/10.5281/zenodo.23426628}, points to the latest archived version as the software develops. The JFK data subset is derived from \texttt{nycflights13::weather} \cite{nycflights13}. Readers who need broader inferential workflows can also turn to established \textsf{R} packages such as \texttt{circular} and \texttt{CircStats}.

\noindent\textbf{Questions for readers.} Can a small mean resultant length arise from two tight clusters? Which formulas are unchanged if $360^\circ$ is added to an angle? When should $\mathbf x$ and $-\mathbf x$ be treated as the same observation? What scientific information can disappear after normalization?

\section{Looking ahead: emerging questions}
Directional statistics has mature foundations, but new data sources keep changing the scale of the questions. Broad frontiers include \emph{high-dimensional directional data}, where normalized representations may contain hundreds or thousands of coordinates; \emph{dynamic directional data}, where orientations evolve over time; \emph{multimodal populations}, where one preferred direction is inadequate; and \emph{uncertainty-aware representations}, where confidence itself must be expressed geometrically. These are open problem areas rather than prescriptions for particular models.

\section{A geometry-first checklist and conclusion}
Before choosing a method, ask: (1) Is the variable periodic? (2) Is the observation a direction or an unoriented axis? (3) Does magnitude itself carry scientific information? (4) Is the sample space a circle, sphere, axis, or higher-dimensional unit sphere? (5) Is there one preferred direction or several? (6) Does the fitted model resemble the observed shape? (7) Would moving an arbitrary angular cut change the conclusion?

These questions capture the article's central message. Familiar statistical ideas do not disappear on circles and spheres; they must be reinterpreted to respect the geometry and symmetries of the sample space. An average becomes a mean direction, variability becomes concentration, probability models wrap around or live on a sphere, axial measurements identify opposite vectors, and cosine similarity reveals spherical geometry inside modern machine learning. Directional statistics is not a replacement for ordinary statistics; it is a reminder to ask, “What mathematical object was measured?” before choosing the statistic.

\noindent\textbf{Data and illustration note.} The JFK wind observations are an extract from the publicly available \texttt{nycflights13} weather data, and the circular statistics are calculated from those observations. The MRI illustration and the document-embedding vectors are pedagogical schematics, not patient data or outputs from a trained language model. The companion R scripts reproduce the underlying teaching examples; the publication-layout illustrations were prepared separately.\par

\section*{Suggested Reading}
Readers who would like to explore these ideas further can begin with Mardia and Jupp's textbook \cite{mardiajupp}, which develops the geometry and probability theory of directional data, and the accessible overview by Pewsey and Garc\'ia-Portugu\'es \cite{pewsey}. For biological applications, Landler, Ruxton, and Malkemper \cite{landler} discuss practical choices in circular-data analysis; the diffusion-MRI references \cite{basser,jones,garyfallidis} explain how orientation arises from medical-imaging measurements. The connection to clustering and machine learning is developed in \cite{banerjee,sra}. To reproduce the wind example, see the \texttt{nycflights13} package \cite{nycflights13}; the accompanying \emph{Directional Statistics Lab} \cite{directionallab} supplies small worked examples and an archived, versioned code release.

\bibliographystyle{unsrt}
\bibliography{Resonance_References_Verified}


\end{document}